\documentclass[twocolumn,aps,prl,epsf]{revtex4}

\usepackage{exscale}                  
\usepackage[intlimits]{amsmath}       
\usepackage{amsfonts}
\usepackage{amssymb,amscd}
\usepackage[dvips]{epsfig}                   
\usepackage{array}
\usepackage[usenames]{color}
\usepackage{graphicx}
\usepackage{bm}
\def\R{\textcolor{black}}

\usepackage{graphics}
\usepackage{graphicx}
\usepackage{dcolumn}
\newcommand{\bs}{\boldsymbol}
\newcommand{\ar}{\arrowvert}
\newcommand{\ra}{\rangle}
\newcommand{\la}{\langle}
\newcommand{\da}{\dagger}

\newcommand{\cd}{\! \cdot \!}
\newcommand{\be}{\begin{equation}}
\newcommand{\ee}{\end{equation}}
\newcommand{\bea}{\begin{eqnarray}}
\newcommand{\eea}{\end{eqnarray}}
 
\begin{document}

\title{
Probing the \R{infrared quark mass from highly excited baryons}
}
\author{P. Bicudo$ ^1$, M. Cardoso$ ^1$, T. Van Cauteren$ ^2$, Felipe J. Llanes-Estrada$ ^3$}
\affiliation{
$ ^1$Instituto Superior T{\'e}cnico, Lisboa, Portugal;  \
$ ^2$Dept. Sub. and Rad. Phys., Ghent University, Belgium; 
$ ^3$Dept. F\'{\i}sica Teorica I, Univ. Complutense, Madrid, Spain}

\begin{abstract}
\R{We argue that three-quark excited states naturally group into
quartets, split into two parity doublets, and that the mass splittings
between these parity partners decrease higher up in the baryon
spectrum. This decreasing mass difference can be used to probe the
running quark mass in the mid-infrared power-law regime. A
measurement of masses of high-partial wave $\Delta^*$ resonances
should be sufficient to unambiguously establish the approximate
degeneracy.  We test this concept} with the first computation of excited
high-j baryon masses in a chirally invariant quark model.
\end{abstract}

\maketitle

Quantum Chromodynamics (QCD) has been thoroughly tested in high-energy
physics, through hadron jets, Drell-Yan processes, $e^-e^+$
annihilation, and deep inelastic scattering. Low-energy QCD manifests
chiral symmetry breaking ($\chi$SB) that enhances quark masses in the
infrared (IR), generating most of the visible mass in our universe, and that
removes the degeneracy from the ground states of light hadron spectra
(so light hadrons do not come in parity doublets). 
\R{ The excited
light-quark baryon spectrum, with masses above that of the
$\Delta$-resonance ($1232$~MeV), has not been accessible}
 to any form
of perturbation theory, effective or fundamental.

Insensitivity to chiral symmetry breaking, recently stressed by
Glozman \cite{Glozman:1999tk,swanson,Wagenbrunn:2006cs}, and
retrospectively present in the excited meson spectra computed in
chirally invariant quark models \cite{Le Yaouanc:1984dr,Bicudo:1998bz}
has led many hadron physicists to accept that spontaneous $\chi$SB,
the salient feature of \R{the low hadron spectrum, actually becomes less
important for highly excited resonances. For these, chiral symmetry is}
asymptotically realized in Wigner mode, arranging hadrons in
degenerate chiral multiplets \cite{Detar:1988kn,Cohen:2001gb}.  
We have realized
that a formalization of this statement is that the ratio of the quark mass
to its momentum $ \langle {m \over k} \rangle $ provides
a new perturbative parameter to study some aspects \R{of the excited
hadron spectrum when the typical momentum becomes higher than
the running mass.} Reasoning in reverse, we propose employing a
cancellation in mass differences of very excited resonances, to access
this small parameter, and hence the quark mass.  We also show
analytically and numerically that three-quark states naturally group
into quartets with two states of each parity.  Diagonalizing the
chiral charge in terms of quarks, the quartet is split into two parity
doublets, \R{and all mass splittings tend to decrease when going higher
in the spectrum.}  For simplicity and to minimize the impact of
molecular meson-nucleon configurations \cite{oset}, we study the
maximum-spin excitations $\Delta^*$ of the Delta baryon.

\begin{figure}[ht!]
\vspace*{-15pt}
\centerline{\includegraphics[height=2.1in]{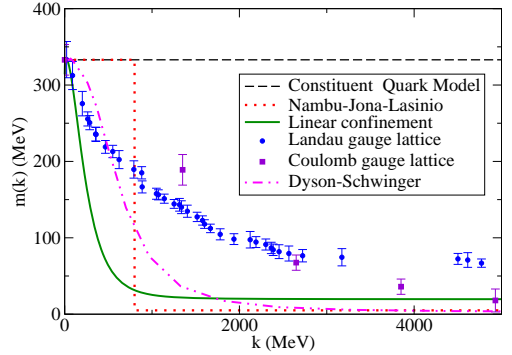}}
\caption{\R{IR enhancement of the light quark mass due to spontaneous
  $\chi$SB}. Shown are quark masses in the main approaches to QCD, all
  multiplied by an arbitrary factor to match them at \R{quark momentum norm $k \to
  0$}. (References in the text.)
\label{fig:massgeneration}}
\vspace*{-15pt}
\end{figure}
The \R{$u$ or $d$ quark mass \cite{pdg}} (understood as taken from the full two-point function
of the field theory), \R{is theorized to vary from $m\simeq 1-6$ MeV at a
high, perturbative momentum $k$ scale (typically \R{$\gg
\Lambda_{\textrm{QCD}} \sim 210$ MeV}), to a constituent mass of circa $300$~MeV at
vanishing momentum \R{$k \simeq \Lambda_{QCD}$}. This IR enhancement of two orders of magnitude
is shown} in Fig. \ref{fig:massgeneration}.  The simplest constituent
quark model considers a constant mass \cite{Paco}, without the
characteristic step-function of the Nambu and Jona-Lasinio model (also
shown).  Breaking chiral symmetry with a linear potential \R{forces a
power-law $m \to {C \over k^4}$ for $m(k) \ll k$. 
Lattice QCD \cite{Bowman:2006zk} and Schwinger-Dyson equations 
\cite{SchwingerDysonLandaugauge}, 
both Euclidian approaches to QCD, also show power-law decays of
the light quark mass. However little has been said about this experimentally, for $k \sim 1.5$ GeV, where we already have $m(k) \ll k$.}

In Hamiltonian dynamics, the quark mass appears in the Dirac spinors
$U_{\kappa\lambda}$ that govern the quark-quark interaction. In the
heavy quark limit, spin-tensor potentials have been successfully
derived by expanding in orders of $k/m(k)$.  For \R{light} quarks the
opposite, \R{ultrarelativistic limit of $k \gg m(k)$} is relevant.
Spinors are then conveniently expanded in the inverse ratio $m(k)/k$,
or \bea \!  \!  \!  \!  \!  U_{\kappa\lambda} = \frac{1}{\sqrt{2E(k)}}
\left[ \begin{array}{c} \sqrt{E(k)+ m(k) }\chi_\lambda \\ \sqrt{E(k)-
      m(k)} \vec{\sigma}\cdot\hat{\kappa} \chi_\lambda\end{array}
    \right] \\ \nonumber \mathop {\longrightarrow }\limits_{k \to
  \infty } \frac{1}{\sqrt{2}} \left[ \begin{array}{c} \chi_\lambda
    \\ \vec{\sigma}\cdot\hat{\kappa} \chi_\lambda
\end{array} \right]
+
 \frac{1}{2 \sqrt{2}} {m(k) \over k} \left[ \begin{array}{c}
 \chi_\lambda \\
-
    \vec{\sigma}\cdot\hat{\kappa} \chi_\lambda
\end{array} \right] \ \ \ \  
\label{msk_expansion}
\eea with $E(k)=\sqrt{k^2+m(k)^2}$, having kept the leading chirally
invariant term, and a next order chiral symmetry breaking $m(k) \over
k$ term. Non-chiral, spin-dependent potentials in the quark-quark
interaction \R{originate from the second term in an expansion of
$H^{\textrm{QCD}}$}
\cite{QCD hamiltonian}
in the weak sense (that is, not of the Hamiltonian operator
itself, but a restriction thereof to the Hilbert space of highly
excited resonances, where $\langle k\rangle$ is large):
\be \label{QCDexp} \la n \ar H^{QCD} \ar n' \ra \simeq \la n \ar
H^{QCD}_\chi \ar n'\ra + \la n \ar \frac{m(k)}{k} H^{QCD\ '}_\chi \ar
n' \ra + \dots \ee
To access the quark mass from experiment one can exploit the
approximate parity degeneracy, that would be exact in the presence of
only the first term $H^{QCD}_\chi$, of excited baryons.  This
degeneracy follows from invariance under transformations generated by
the chiral charge~\cite{Nefediev:2008dv} $ Q_5^a=\int d{\bf x}
\psi^\dagger (x) \gamma_5 \frac{\tau^a}{2} \psi (x) $ and
$[Q_5^a,H]=0$. However Chiral Symmetry is spontaneously broken by the
ground state, $Q_5^a \ar 0 \ra \not = 0$ providing a large quark mass
in the propagator, $m(k)$, pseudo-Goldstone bosons ($\pi,\ K,\ \eta$),
and the loss of parity degeneracy in ground state baryons.
Substituting the spinors, and  in terms of Bogoliubov-rotated $q\bar{q}$ 
\R{normal modes $B$ and $D$, $Q_5$ becomes}
\begin{eqnarray} \label{chiralcharge}
Q_5^a  = \int \frac{d^3k}{(2\pi)^3} \sum_{\lambda
\lambda ' f f'c} \left(  \frac{\tau^a}{2} \right)_{ff'}
{ k \over \sqrt{ k^2 + m^2(k)}}
 \\ \nonumber 
\left( ({\boldsymbol \sigma}\cd{\bf
\hat{k}})_{\lambda \lambda'}  	
\left( B^\da_{k \lambda f c} B_{k \lambda' f' c} + D^\da_{-k \lambda' f' c}
D_{-k \lambda f c}
\right) + \right. \\  \nonumber \left.
{ m(k) \over k} (i\sigma_2)_{\lambda \lambda'} \
\left( B^{\da}_{k\lambda f c} D^\da_{-k\lambda'f'c}+
B_{k \lambda' f' c} D_{-k \lambda f c}
\right) \right) \ .
\end{eqnarray}
In the presence of Spontaneous $\chi$SB, $ m(k) \not=0$, and the last
term realizes chiral symmetry non-linearly in the spectrum as it
creates/destroys a pion.
\R{But when \R{$\la k \ra \gg m(k)$}, the ${\boldsymbol \sigma}\cd{\bf \hat{k}}$-term
dominates and chiral symmetry is realized linearly (with only quark
counting operators flipping parity and spin).  This happens for a
highly excited baryon resonance whose constituents have a momentum
distribution peaked at higher momenta than the IR enhancement of
$m(k)$ (Fig.~\ref{fig:massgeneration}).}

For the ground--state resonances in excited spin channels,
three--quark variational wavefunctions are relevant and a reasonable
phenomenological guide. When $Q_5$ acts on such wavefunction \R{$\ar N
\ra =F_{ijk} B^\da_i B^\da_j B^\da_k \ar \Omega \ra$}, the result also
contains three quarks, but one of them is spin-rotated from
$B_{k\lambda}$ to $\sigma\cd\hat{k}_{\lambda
  \lambda'}B_{k\lambda'}$. Successive application of the chiral charge
spin-rotates further quarks, changing each time the parity of the
total wavefunction.  However the sequence of states is closed since
$\sigma\cd\hat{k}\sigma\cd\hat{k}={\mathbb{I}}$. In fact, starting
with an arbitrary spin--symmetric wavefunction with parity $P$, one
generates our claimed Baryon Quartet of equal isospin (\R{isospin} index
dropped):
\begin{eqnarray}
\ar N_0^P \ra = \sum F^P_{ijk} B^\da_i B^\da_j B^\da_k \ar \Omega \ra 
\\ \nonumber
\ar N_1^{-P} \ra = \frac{1}{3}\sum F^P_{ijk}  \left(
\left({\boldsymbol \sigma}\cd{\bf \hat{k}} B^\da\right)_i B^\da_j B^\da_k +
\right. \\ \nonumber  \left.
B^\da_i \left({\boldsymbol \sigma}\cd{\bf \hat{k}} B^\da\right)_j B^\da_k  +
 B^\da_i B^\da_j \left({\boldsymbol \sigma}\cd{\bf \hat{k}} B^\da\right)_k 
\right) \ar \Omega \ra 
\end{eqnarray}
\begin{eqnarray} \nonumber
\ar N_2^{P} \ra = \frac{1}{3} \sum F^P_{ijk} 
\left(
\left({\boldsymbol \sigma}\cd {\bf \hat{k}} B^\da\right)_i
\left({\boldsymbol \sigma}\cd {\bf \hat{k}} B^\da\right)_j 
B^\da_k 
+ \right. \\ \nonumber \left.
B^\da_i
\left({\boldsymbol \sigma}\cd {\bf \hat{k}} B^\da\right)_j
\left({\boldsymbol \sigma}\cd {\bf \hat{k}} B^\da\right)_k  
+
\left({\boldsymbol \sigma}\cd {\bf \hat{k}} B^\da\right)_i
B^\da_j
\left({\boldsymbol \sigma}\cd {\bf \hat{k}} B^\da\right)_k 
\right) \ar \Omega \ra 
\\ \nonumber
\ar N_3^{-P} \ra = \sum F^P_{ijk} \left({\boldsymbol \sigma}\cd
{\bf \hat{k}} B^\da\right)_i
\left({\boldsymbol \sigma}\cd {\bf \hat{k}} B^\da\right)_j 
\left({\boldsymbol \sigma}\cd {\bf \hat{k}} B^\da\right)_k \ar \Omega \ra
\end{eqnarray}
which is a natural basis to discuss chiral symmetry restoration in
baryons, through wavefunctions that are linear combinations $\ar N \ra
= \sum c_a \ar N_a \ra$.
\R{This reducible representation of $Q_5^{a=3}$ corresponds to $[(3/2,0) \oplus (0,3/2)]^2$ of $SU(2)_L \times SU(2)_R$, and can be embedded in larger ones including nucleons
\cite{Cohen:2001gb}
.}

Because $[Q_5,H]=0$, the two can be diagonalized simultaneously. The
representation of the chiral charge in the (non-orthonormal) quartet
coordinates breaks in two blocks since $Q_5$ changes parity, or using
the square charge \be Q_5^{ 2} \left( \begin{tabular}{c} $c_0$
  \\ $c_2$ \end{tabular} \right) =\left[ \begin{tabular}{cc} 3& 2\\ 6
    &7
\end{tabular}\right] \left( \begin{tabular}{c}
$c_0$ \\ $c_2$ \end{tabular} \right) \ .  \ee Immediately one sees
that the two linear combinations $ N_0-N_2 $ and $N_0+3N_2$
diagonalize the square chiral charge in the positive parity sector (if
the original wavefunction had positive parity), with $ N_1-N_3 $ and
$3N_1+N_3$ doing so for negative parity.
The quartet then separates into two doublets connected by the chiral
charge \bea Q_5 (N_0 - N_2) = N_1-N_3 & Q_5 (N_0+3N_2) &= 3(3N_1+N_3)
\nonumber \\ Q_5 (N_1 - N_3) = N_0-N_2 & Q_5 (3N_1+N_3) &=
3(N_0+3N_2)\ .  \eea Two doublets appear with different eigenvalues of
$Q_5^2$, 1 and 9 respectively, and the interdoublet splitting becomes
a dynamical question (we will argue shortly that it is small for
highly excited baryons). However, the splitting within the doublet
\emph{must vanish} asymptotically.  Even for fixed (not running) quark
mass, when the typical kinetic energy is high enough $\la k \ra >> m$,
the effects of the quark mass are negligible. Parity doubling then
\R{boils down} to whether the interaction terms are also chiral symmetry
violating or not.


\begin{figure}[t!]
\centerline{\includegraphics[width=3in,height=1.9in]{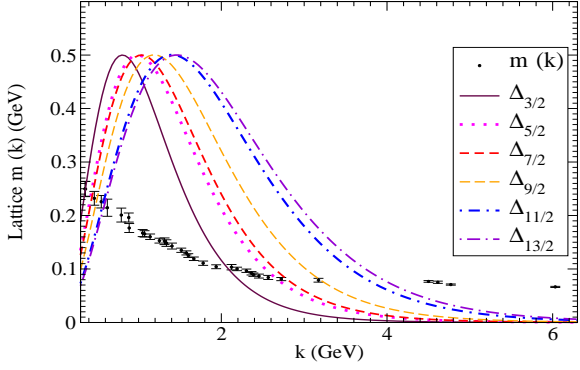}}
\caption{Typical momentum distributions of increasingly excited
  $\Delta_{3/2}$, \dots $\Delta_{13/2}$ resonances overlap less and
  less with the dynamically generated IR quark mass so that $\la n\ar
  \frac{m(k)}{k} H^{QCD\ '}_\chi \ar n \ra$ becomes
  small. (Illustrative variational wavefunctions for a linear
  potential with string tension $\sigma=0.135$ GeV$^2$, not normalized
  for visibility).\label{fig:anglinmoment}}
\vspace*{-12pt}
\end{figure}

\hspace{0.1 cm}

We now exploit the smallness of the $| M^{P=+} - M^{P=-}| $ mass
difference with increasing angular momentum to guide experiment in
obtaining information about the running quark mass. For this we
establish how the \R{$j$-dependence} of the splitting is related to the
\R{$k$-dependence} of $m(k)$.
The $M^\pm$ in our proposed study are the
masses of the ground state quartets of the $\Delta$ spectrum, with
parity $\pm$ and in the limit of large total angular momentum
$j>>3/2$.  The two approximately degenerate masses $M^+$ and $M^-$ are
in the same leading linear Regge trajectory, phenomenologically fixing
their \R{$j$-dependence} to 
\be \label{trajectory} 
j = \alpha_0 + \alpha
{M^\pm}^2
\mathop {\longrightarrow }\limits_{j \to \infty }
\alpha {M^\pm}^2 \ .  
\ee 
The parity of the ground state alternates
between $+$ and $-$ as the angular momentum steps up by one.  Large
$j$ is equivalent to large quark orbital angular momentum \R{$l \gg 1$ since
the spin is finite, and also to a large average linear momentum
$\langle k \rangle \gg \Lambda_{QCD}$. Figure \ref{fig:anglinmoment} shows how the
wavefunctions of high-$j$ states \R{overlap} with the running
$m(k)$. In these states $\la m(k) /k \ra$ is a useful perturbative parameter. }From the relativistic virial theorem~\cite{virial} between
kinetic and total energies, 
\be 
\label{Regge} \langle k \rangle \to
c_2 \, M^\pm \to {c_2 \over \sqrt \alpha} \sqrt j \ , 
\ee 
(where $ c
_2$ is a constant; for \R{instance in the case of } a linear potential model, and 3 quarks in a
baryon, $c_2=1/6$).

The first term in Eq. (\ref{QCDexp}) cancels out in the difference $
\ar M^+ - M^-\ar << M^\pm $ (while both $M^\pm$ are dominated by the
chirally invariant term, their difference stems from the dynamically
generated quark mass) thus exposing the second term in
eq.(\ref{QCDexp}), proportional to $ \langle {m(k) \over k} \rangle$,
viz.  \bea \label{paritysplitofm}\! \! \!  \ar M^+\! -\! M^- \ar \!
\to \!  \langle { m( k )\over k } H^{QCD\ '}_{\chi} \rangle \to c_3{ m
  ( \langle k \rangle ) \over \langle k \rangle } \langle
H^{QCD\ '}_{\chi} \rangle \eea (the factorization is allowed by the
mean-value theorem at the price of a constant that we do not attempt
to determine here).  This equation is analogous to the renowned
Gell-Mann-Oakes-Renner relation $ M_\pi^2 = - m_q \frac{\la \bar{\psi}
  \psi \ra}{f_\pi^2} $ but active when chiral symmetry is realized
linearly, as in the high-baryon excitations we examine.


\R{Notice that $H^{QCD\ '}_{\chi}$ differs from $H^{QCD}_{\chi}$
due to the $-$ sign of the $({\boldsymbol \sigma} \cd \ {\bf \hat{k}})_i$ term in the $\la m(k) /k \ra$ expansion of the quark $i$ spinor
in eq. (\ref{msk_expansion}). This changes the sign of spin-dependent
terms in $H^{QCD\ '}_{\chi}$, led in high $j$ by the spin-orbit term.
 In  $H^{QCD}_{\chi}$ the spin-orbit is crucial to correct the 
angular momentum in the centrifugal barrier term from 
${{\bf L}_i }^2$ to the chirally invariant $ {{\bf L}_i}^2 +
2 {\bf L}_i \cdot {\bf S}_i = {{\bf J}_i}^2 - {3 \over 4}$.
But the opposite sign of the spin-orbit term of $H^{QCD\ '}_{\chi}$ 
produces a $\ar M^+ - M^-\ar$ mass difference, enhanced in the IR by 
$\la m(k) /k \ra$.
Since the centrifugal barrier scales in the large $j$ limit like the 
baryon mass $M^\pm$,
and the spin-orbit term comes with one power of $j$ less than it,
\be
\langle H^{QCD\ '}_{\chi} \rangle \to c_5 \, M^\pm j^{-1} \to {c_5
  \over \sqrt \alpha} \sqrt {1 \over j} \ .
\label{tensor}
\ee
Combining the $j$-power-law analysis of eqs. (\ref{trajectory}) 
to (\ref{tensor}), 
\bea \! \!  \ar
M^+ - M^-\ar\!  \to \!  {c_3 c_5 \over
  c_2 \sqrt \alpha} m ( \langle k \rangle ) \, j ^{-1}  
\label{total scaling}
\eea 
that links the IR
enhancement of the quark mass to baryon spectroscopy in a usable way.
An experimental extraction proceeds by just fitting the exponent $-i$ of
$j$ in the splitting $\ar M^+ - M^- \ar \propto j^{-i} $.  Then, we obtain, 
\bea
m(\frac{c_2}{\sqrt{\alpha}} \times \sqrt{j}) \propto j^{-i+1} \ \ {\rm
  and} \ \ m(k) \propto k^{-2i+2} \ .  
\label{final j scaling}  
\eea 
The same exponent $i$ in eq. (\ref{final j scaling})
}can be obtained from the fit 
\cite{For a linear...}
to the $\ar M^+ - M^- \ar$
with increasing $j$! 
Though the quark mass itself is gauge--dependent, our
analysis suggests that its power--law exponent is not, since it is
directly related to an observable.


\begin{figure}[t!]
\centerline{\includegraphics[width=2.7in]{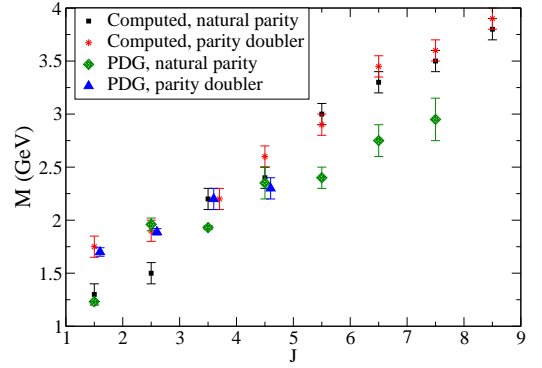}}
\caption{\R{Parity doubling in the spin-excited $\Delta$ spectrum for a
  string tension $\sigma=0.135$~GeV$^2$ and the mass gap angle matched
  to the lattice data. } A $qqq$ variational \R{Monte-Carlo} computation of
  the masses with a chiral Hamiltonian shows that the ground-states
  with parity $\pm$ for each $j=3/2\dots 13/2$ quickly
  degenerate. Experimentally the degeneracy can be claimed for the
  $9/2$ states only.
\label{fig:doubling}}
\vspace*{-12pt}
\end{figure}

We next contribute the first competitive chirally invariant quark
model computation of parity doubling in the excited baryon spectrum,
shown in figure \ref{fig:doubling}, employing a well-known model of
Coulomb-QCD that can be seen as a field theory upgrade of the Cornell
model, \bea \nonumber H = - g_s \int d{\bf x} \Psi^\da (x) \alpha \cd
{\bf A}(x) \Psi (x) + Tr \int d{\bf x} ( {\bf E} \cd {\bf E} + {\bf B}
\cd {\bf B} ) \\ \nonumber +
\int d {\bf x} \Psi ^{\dagger} _q
 (-i
\alpha\cd \nabla + \beta m ) \Psi_q \!  -\! \frac{1}{2}\! \int\! d\!
          {\bf x} d\! {\bf y} \rho^a_{\bf x}V_L(\arrowvert {\bf x} -
          {\bf y} \arrowvert) \rho^a_{\bf y} \eea with the kernel
          $V_L(r)=\sigma r$ with string tension $\sigma=0.135$
          GeV$^2$, coupled to the color charge density $ \rho^a({\bs
            x}) = \Psi^\dagger({\bs x}) T^a\Psi({\bs x}) +f^{abc}{\bf
            A}^b({\bs x})\cdot{\bf \Pi}^c({\bs x}) \ .  $ We solve the
          BCS gap equation and employ both the so calculated $m(k)$
          quark mass as well as lattice computation in Landau and
          Coulomb gauges \cite{Bowman:2006zk}.  The model has the same
          chiral structure of QCD\cite{Le Yaouanc:1984dr}, satisfying
          the Gell-Mann-Oakes-Renner relation and other low-energy
          theorems, and allowing computations of static $N\pi$
          observables \cite{Bicudo:2001cg}.
\R{ Pauli antisymmetrization
          of wavefunctions, spin summations and the 9-dimensional
          integrations are performed in a computer program, employing
          \R{Monte-Carlo} integration~\cite{Hahn:2004fe}, to compute the
          excited $\Delta^*$ spectrum.} We proceed variationally and
          employ several types of wavefunctions.  While a suboptimal
          variational basis for low-lying states, the quartet states
          coincide with $H$-eigenstates high in the spectrum. 
\R{We use
          both the quartet and the fixed $L$ variational basis.}

\begin{figure}[t!]
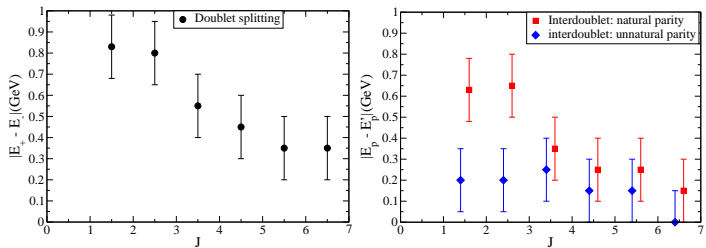

\centerline{\includegraphics[width=1.8in]{split1.eps}
\includegraphics[width=1.8in]{split2.eps}
}
\caption{\R{All model quartet splittings decrease for high $j$. Left:
  \textit{intradoublet} splitting for the first parity doublet. Right:
  \textit{interdoublet} splittings between natural and unnatural
  parity states.  The latter is smaller for low $j$.}
\label{fig:splittings}}
\vspace*{-12pt}
\end{figure}

As can be seen from figure \ref{fig:splittings}, within our
variational and \R{Monte-Carlo} integration errors, the model splittings
drop with $j$, as predicted analytically; a model statement equivalent
to the generic eq. (\ref{paritysplitofm}) reads \bea \! \!  \! \!
M_+-M_- = 3 \int \frac{d^3k_1}{(2\pi)^3} \frac{d^3k_2}{(2\pi)^3}
\left(\frac{2}{3}\right)\int \frac{d^3q}{(2\pi)^3} \\ \nonumber
\hat{V}(q) \frac{1}{2} \left( \frac{m(\ar {\bf k}_1\ar)}{\ar {\bf
    k}_1\ar} + \frac{m(\ar {\bf k}_1+ {\bf q}\ar)}{\ar {\bf k}_1+ {\bf
    q}\ar} \right) F^{*\lambda_1\lambda_2\lambda_3} ({\bf k}_1, {\bf
  k}_2) \\ \nonumber
\left( 
\mathbb{I} - 
{\boldsymbol \sigma}\hat{\bf k}_1 
 {\boldsymbol \sigma}
\widehat{{\bf k}_1+{\bf q}}
\right)_{\lambda_1 \mu_1} F^{\mu_1\lambda_2\lambda_3}
({\bf k}_1+{\bf q}, {\bf k}_2-{\bf q}  )
\eea  

If precise data becomes available at ELSA or Jefferson Lab for the
$\Delta_{J}$ with $J=7/2,9/2,11/2, 13/2 \cdots$ parity quartets, one
should be able to distinguish between the $1/j$ fall of
$|M^+-M^-|$ for non-chiral models (with a constant difference
$(M^+-M^-)^2$ between the Regge Trajectories \cite{Paco}), and the
faster drops for chiral theories such as QCD.  Since the two doublets
are closely degenerate, both positive and negative parity ground
states will have a nearby resonance with identical quantum numbers.
Given their width, it is likely they will only be distinguished by
very careful exclusive decay analysis.

Further, if the $\Delta$ spectrum could be measured up to high
resonance masses, beyond the IR $m(k)$ enhancement, and a lattice
calculation of $\la H^{QCD\ '}_{\chi}\ra$ became available, an almost
direct measurement of the current quark mass would follow in a regime
of logarithmic, rather than power-law running. 
\R{We advocate that a
measurement of masses of high partial-wave $\Delta$ resonances with an
accuracy of 50 MeV should suffice to establish the approximate
degeneracy and test the concept of running quark mass in the IR. The
presented concepts should also further motivate analysis of empirical
data\cite{JuliaDiaz:2007fa} such as EBAC (Excited Baryon Analysis
Center) at Jefferson Lab as well as Lattice QCD collaborations who are
addressing the baryon spectrum.}

\begin{acknowledgments}
We thank L. Glozman for useful conversations, and grants FPA
2008-00592/FPA, FIS2008-01323, CERN/FP /83582/2008, POCI/FP
/81933/2007, /81913/2007 , PDCT/FP /63907/2005 and /63923/2005,
Spain-Portugal billateral grant HP2006-0018 / E-56/07, as well as the
Scientific Research Fund of Flanders.
\end{acknowledgments}


\begin{thebibliography}{99}

\bibitem{Glozman:1999tk}
  L.~Y.~Glozman,
  Phys.\ Lett.\  B {\bf 475} (2000) 329.

\bibitem{swanson}
  E.~S.~Swanson,
  Phys.\ Lett.\  B {\bf 582}, 167 (2004).

\bibitem{Wagenbrunn:2006cs}
  R.~F.~Wagenbrunn and L.~Y.~Glozman,
  Phys.\ Lett.\  B {\bf 643}, 98 (2006);
  T.~D.~Cohen and L.~Y.~Glozman,
  Mod.\ Phys.\ Lett.\  A {\bf 21}, 1939 (2006);
  L.~Y.~Glozman, A.~V.~Nefediev and J.~E.~F.~Ribeiro,
  Phys.\ Rev.\  D {\bf 72}, 094002 (2005).



\bibitem{Le Yaouanc:1984dr}
  A.~Le Yaouanc {\it et al.},
  Phys.\ Rev.\  D {\bf 31} (1985) 137.
  
\bibitem{Bicudo:1998bz}
  P.~Bicudo {\it et al.},
  Phys.\ Lett.\  B {\bf 442}, 349 (1998).

\bibitem{Detar:1988kn}
  C.~E.~DeTar and T.~Kunihiro,
  Phys.\ Rev.\  D {\bf 39}, 2805 (1989);
  D.~Jido, T.~Hatsuda and T.~Kunihiro,
  Phys.\ Rev.\ Lett.\  {\bf 84}, 3252 (2000).

\bibitem{Cohen:2001gb}
\R{
  T.~D.~Cohen and L.~Y.~Glozman,
  Phys.\ Rev.\  D {\bf 65}, 016006 (2001)
  [arXiv:hep-ph/0102206].
}


\bibitem{oset}
  S.~Sarkar {\it et al.},
  arXiv:0902.3150 [hep-ph].
  

\bibitem{pdg}
  C.~Amsler {\it et al.}  [PDG],
  Phys.\ Lett.\  B {\bf 667}, 1 (2008).

\bibitem{Paco}
  J.~Segovia, D.~R.~Entem and F.~Fernandez,
  Phys.\ Lett.\  B {\bf 662}, 33 (2008).

  
  

\bibitem{Bowman:2006zk}
P.~O.~Bowman, {\it et al.},
  Nucl.\ Phys.\ Proc.\ Suppl.\  {\bf 161}, 27 (2006);
  M.~B.~Parappilly {\it et al},
  Phys.\ Rev.\  D {\bf 73}, 054504 (2006).
  S.~Furui,
  arXiv:0801.0325 [hep-lat].
  
  
\bibitem{SchwingerDysonLandaugauge}
  R.~Alkofer {\it et al.}, Ann. Phys. NY (in press)
  arXiv:0804.3042 [hep-ph].




  
\bibitem{QCD hamiltonian}
\R{For a discussion on $H^{\textrm{QCD}}$, a good starting
  point is N. H. Christ and T. D. Lee, Phys. Rev. D\textbf{22} (1980)
  939.}

\bibitem{Nefediev:2008dv}
  A.~V.~Nefediev, J.~E.~F.~Ribeiro and A.~P.~Szczepaniak,
  JETP Lett.\  {\bf 87}, 271 (2008).

 
\bibitem{virial}
  W.~Lucha and F.~F.~Schoberl,
  Mod.\ Phys.\ Lett.\  A {\bf 5}, 2473 (1990).


\bibitem{For a linear...} 
For a linear potential, with
  $m(k) \propto k^{-4}$, $\ar M^+ - M^- \ar \propto {1 \over j^3}$.  If,
  on the contrary, the quark mass is constant, and still the potential
  remains \R{chirally} invariant, then the decrease follows a slower $ j
  ^{-1}$.  We can retrospectively understand analytically the
  $q\bar{q}$ numerical results of \cite{Wagenbrunn:2006cs}, with
  chiral quartets and with the $j$-scaling of the
  splittings. \R{While their doublet $\ar M^+ - M^- \ar \sim {1 \over j^3}$;
  their interdoublet splitting follows a weaker 
  $1 \over j^{3 / 2}$ as expected in a corollary of eq. (\ref{tensor})     
  for spin-independent potentials}. 


\bibitem{Bicudo:2001cg}
  P.~Bicudo, G.~Krein and J.~E.~F.~Ribeiro,
  Phys.\ Rev.\  C {\bf 64} (2001) 025202.


\bibitem{Hahn:2004fe}
  T.~Hahn,
  Comput.\ Phys.\ Commun.\  {\bf 168}, 78 (2005).


\bibitem{JuliaDiaz:2007fa}
  B.~Julia-Diaz {\it et al},
  Phys.\ Rev.\  C {\bf 77} (2008) 045205.



\end{thebibliography}
\end{document}